\documentclass[aps,showpacs,preprint,epsf,epsfig, nofootinbib,titlepage]{revtex4-1}
\usepackage{graphicx}
\usepackage{dcolumn}
\usepackage{bm}
\usepackage{hyperref}
\usepackage{epsfig,graphicx,color,appendix}
\usepackage{multirow}
\usepackage{array}
\usepackage{tabularx}
\usepackage{capt-of}
\usepackage{amsmath}
\usepackage{caption}
\usepackage[countmax]{subfloat}

\usepackage{enumitem}
\setlist{nolistsep}
\usepackage{float}
\usepackage{booktabs}
\usepackage[absolute]{textpos}
\textblockorigin{10mm}{10mm} 
\begin{document}
\def\bibsection{\section*{\refname}} 
\def\|{\vert}
\def\ol{\bar}
\def\ov{\bar}
\def\be{\begin{eqnarray}}
\def\en{\end{eqnarray}}
\def\non{\nonumber}
\def\t{\times}
\def\lb{\Lambda}
\def\lbc{\Lambda_c^{+}}
\def\om{\Omega}
\def\omc{\Omega_c}
\def\oc{\Omega_c^0}
\def\occ{\Omega_{cc}^{+}}
\def\oc{\Omega_c^0}
\def\xic{\Xi_c}
\def\xicd{\Xi_c^{'}}
\def\xicc{\Xi_{cc}^{++}}
\def\xicp{\Xi_{cc}^{+}}
\def\xix{\Xi_{cc}}
\def\x{\Xi}
\def\si{\Sigma}
\def\sic{\Sigma_c}
\def\sicp{\Sigma_c^{++}}
\def\k{{K_1}}
\def\ds{{D_s}}
\def\k1d{\underline{K}_{1}}
\def\kb0{\bar{K}^{0}}
\def\kb{\bar{K}_{1}}
\def\kbd{\underline{\bar K}_{1}}
\def\etap{\eta^{'}}
\def\ra{\rangle}
\def\A{{\cal A}}
\def\B{{\cal B}}
\def\L{{\cal L}}
\def\ben{\begin{eqnarray}}
\def\en{\end{eqnarray}}
\def\non{\nonumber}
\def\la{\langle}
\def\ra{\rangle}
\def\t{\times}
\def\pp{{\prime\prime}}
\def\nc{N_c^{\rm eff}}
\def\vp{\varepsilon}
\def\hep{\hat{\varepsilon}}
\def\up{\uparrow}
\def\dw{\downarrow}
\def\vma{{_{V-A}}}
\def\vpa{{_{V+A}}}
\def\smp{{_{S-P}}}
\def\spp{{_{S+P}}}
\def\J{{J/\psi}}
\def\ov{\bar}
\def\Lqcd{{\Lambda_{\rm QCD}}}
\long\def\symbolfootnote[#1]#2{\begingroup%
\def\thefootnote{\fnsymbol{footnote}}\footnote[#1]{#2}\endgroup}
\def\lsim{ {\ \lower-1.2pt\vbox{\hbox{\rlap{$<$}\lower5pt\vbox{\hbox{$\sim$}
}}}\ } }
\def\gsim{ {\ \lower-1.2pt\vbox{\hbox{\rlap{$>$}\lower5pt\vbox{\hbox{$\sim$}
}}}\ } }

\font\el=cmbx10 scaled \magstep2{\obeylines\hfill \today}

\title{ \Large Estimates of W-exchange contributions to $\Xi_{cc}$ decays}

\author{\bf Neelesh Sharma$^{\ddagger}$  and Rohit Dhir$^{\dagger}$ }
\email{$^{\dagger}$dhir.rohit@gmail.com, $^{\ddagger}${nishu.vats@gmail.com} }

\affiliation{\sl  Research Institute \\and \\Department of Physics and Nanotecnology, SRM University, Chennai 603203, India
\\
\\
\\
\\
}

\begin{abstract}
Encouraged by the recent discovery of
the $\Xi_{cc}$ baryon, we investigate two-body nonleptonic weak decays of doubly charmed, $\Xi_{cc}$, baryons. We calculate the branching ratios for 
Cabibbo-Kobayashi-Maskawa–favored 
and -suppressed modes in factorization and pole model approaches. The preliminary estimates of nonfactorizable W-exchange contributions are obtained using the pole model. We find that the W-exchange contributions to $\Xi_{cc}$ decays, being sizable, cannot be ignored.
\end{abstract}

\maketitle
\thispagestyle{empty}

\setcounter{page}{1}

\section{INTRODUCTION}

 The resolution of the longstanding puzzle involving the $\Xi_{cc}$ states has been much awaited since their first observations were reported by the SELEX collaboration \cite{1,2}. Most recently, the LHCb collaboration announced the observation of a doubly charmed $\xicc$ baryon \cite{3} found in the $\Lambda_c^+ K^-\pi^+\pi^+$ mass spectrum.  The mass of the observed state is determined as $M_{\xicc}=3621.40 \pm 0.72 \pm 0.27 \pm 0.14 {~\rm MeV}$, while the mass difference $M_{\xicc}-M_{\lbc}=1334.94 \pm 0.72 \pm 0.27{~\rm MeV}$. The new observation of a doubly heavy charm baryon has revamped the interest of heavy flavor physicists as being a good candidate with which to  study the heavy-quark dynamics. Although, the life time, $\tau_{\Xi_{cc}}$, has not yet been given experimentally, ample theoretical estimates exist in literature that range from $\sim 50 - 670$ fs \cite{4,5,6,7,8,9}. Another interesting aspect of doubly heavy baryons is their spectroscopy \cite{8,9,10,11}. In addition to the three quark dynamics, the doubly
heavy baryons can be identified by the set of quantum numbers $(J^P,S_d)$ in diquark picture, where $S_d$ is the spin of the heavy diquark. Thus, spins of the two heavy-quarks are coupled to form the ($S_d=1$) symmetric spin configuration of a diquark $\{Q_{1}Q_{2}\}$ and the ($S_d=0$) antisymmetric spin configuration of a diquark $[Q_{1}Q_{2}]$. The general convention is to denote the antisymmetric state as a primed one i.e. $\left| B^{\prime}\right\rangle $  and symmetric heavy-diquark state as an unprimed, $\left| B \right\rangle$ state. Also, the wave functions of the $\left| B \right\rangle$ and $\left| B^{\prime}\right\rangle $ states are expected to mix \cite{12,13,14,15,16,17,18}. However, in the present work we consider three quarks as a independent dynamical entities.

Theoretically, the mass
spectra, magnetic moments, and radiative and semi-leptonic decays of the doubly charmed baryons has been the center of interest for the last decade \cite{4,5,6,7,8,9,10,11,12,13,14,15,16,17,18,19,20,
21,22,23,24,25,26,27,28,29,30,31,32,33,34,35,36}. On the contrary, the progress in the heavy-baryon nonleptonic weak decays has been very slow \cite{37,38,39,40,41,42,43,44,45,46}, although, the recent experimental observations have revived the activities in nonleptonic decays of heavy baryons in last few years \cite{47,48,49,50,51,52,53,54,55,56,57,58,59,60,61}.  Thus, we put our focus on the two-body nonleptonic weak  decays of doubly charmed baryons. Very recently, weak decays of doubly heavy baryons were analyzed in SU(3) symmetry and in the quark-diquark picture using factorization and the light front approach \cite{57,58}. In another interesting work, the analysis of factorizable $\Xi_{cc}^{++} \to \Sigma_c^{++} \bar{K}^{(*)0}$ decays was carried out using the covariant confined quark model (CCQM) \cite{59}. The theoretical interpretation of the experimentally favored decay chain  $\Xi_{cc}^{++} \to \Sigma_c^{++} (\to \Lambda_c^+ \pi^+) + \bar K^{*0} (\to K^-  \pi^+)$ due to the dominant branching ratios of the daughter decays is first presented in \cite{60}. In addition, the short-distance and long-distance (W-exchange) contributions to the decay channels of $\x_{cc}$ baryons are calculated more systematically using factorization and final-state interaction (FSI) rescattering, respectively \cite{60}. The branching ratios of nonleptonic decays of the doubly heavy baryons are predicted in the perturbative QCD (pQCD) \cite{61}. 

It may be emphasized that in heavy baryon decays, unlike meson decays, the W-exchange contributions could be as important as factorizable for being free from helicity and color suppression \cite{62,63,64,65,66,67,68,69,70,71}. In fact, many of the observed charm baryon decays receive contributions solely from W-exchange diagrams. Therefore, in the present work, we give preliminary estimates of W-exchange (pole) contributions using the pole model. To obtain the factorization contributions, we use the nonrelativistic quark model (NRQM)
\cite{72} and heavy quark effective theory (HQET) \cite{73} based form factors, which has worked reasonably well in explaining weak decays of charm baryons. Also, we use a more accurate approach \cite{74} to include SU(4) symmetry-breaking effects in evaluation of meson-baryon strong couplings. To calculate the pole amplitude (W-exchange contributions), we use nonrelativistic approximation \cite{37,75} to evaluate weak matrix element. It may be noted that as a first estimate of pole contributions we consider ground-state $\frac{1}{2}^+-$ intermediate baryon pole terms only. Moreover, the $\frac{1}{2}^--$ intermediate pole terms are difficult (for being nontrivial) to calculate and a little is known about the strong couplings constants of $\frac{1}{2}^-$ baryons involved. It has been shown in past \cite{70,76} that SU(4) symmetry-breaking could be induced by the variation of a spatial baryon wave function overlap in weak decay amplitude. We find that pole contributions are significantly enhanced upon the inclusion of flavor-dependent effects via  $|\psi(0)|^2$ variation, consequently, we get larger branching ratios for the decays involving W-exchange diagrams. Thus, a number of decays have sizable branching ratios that could be suitable for  future experimental measurements at LHCb, CEPC Belle II, etc.

The present paper is organized as follows. In Sec. II, we give the Hamiltonian and decay rate formula. Section III deals with the evaluation of decay amplitudes. Numerical results and discussions are given in Sec. IV. We summarize our findings in the last section.

\section{HAMILTONIAN AND DECAY RATE}

The charm changing two-body nonleptonic decays of (doubly heavy) baryons, emitting pseudoscalar ($P$) meson,  proceed through usual current $\otimes $ current effective weak Hamiltonian,
\ben
{\cal H}_{W}^{\text{eff}}= \frac{G_{F}}{ \sqrt{2}} \Big\{ V_{ud} V_{cs}^{*} &\big[& c_{1} (\bar{u}d)_{V-A}(\bar{c}s)_{V-A}+c_{2} (\bar{s}d)_{V-A}(\bar{u}c)_{V-A}\big]_{(\Delta C = \Delta S = -1)} + \non \\ V_{ud} V_{cd}^{*} &\big[& c_{1} \{(\bar{s}c)_{V-A}(\bar{u}s)_{V-A}-(\bar{d}c)_{V-A}(\bar{u}d)_{V-A}\}  + \non \\  &~~& c_{2}\{ (\bar{u}c)_{V-A}(\bar{s}s)_{V-A}-(\bar{u}c)_{V-A}(\bar{d}d)_{V-A}\}\big]_{(\Delta C = -1,~  \Delta S = 0)} - \non \\ V_{us} V_{cd}^{*} &\big[& c_{1} (\bar{d}c)_{V-A}(\bar{u}s)_{V-A}+c_{2} (\bar{u}c)_{V-A}(\bar{d}s)_{V-A}\big]_{(\Delta C =-\Delta S= -1)}\Big\} ,
\en
where $V_{ij}$ denote the Cabibbo-Kobayashi-Maskawa
(CKM) matrix elements and $(\bar{q}_{i} q_{j} )_{V-A}\equiv \bar{q}_{i} \gamma _{\mu } (1-\gamma _{5} )q_{j} $, the weak \textit{V-A }current. The Hamiltonian consists of CKM-favored ($\Delta C = \Delta S = -1$), CKM-suppressed ($\Delta C = -1,  \Delta S = 0$) and CKM-doubly-suppressed ($\Delta C = - \Delta S = -1$) decay modes. The QCD (Wilson) coefficients  $c_{1} (\mu )=1.2$, $c_{2} (\mu )=-0.51$ at $\mu \approx m_{c}^{2} $ in the large $N_c$ limit are used in analysis (for a review see \cite{77}). The coefficients $c_1$ and $c_2$ may be treated as free parameters for being affected by nonfactorizable contributions. 
In general, the transition amplitude can be expressed in terms of reduced matrix element for $B_{i} (\frac{1}{2}^{+},~ p_i )\to B_{f} (\frac{1}{2}^{+},~ p_f )+P_{k} (0^{-},~q )$ decay process:
\begin{equation}
{\cal A}(B_i\to B_{f}P) \equiv \la B_{f} (p_{f} )P_{k} (q)\vert {\cal H}_{W}^{\text{eff}} \vert B_{i} (p_{i} )\ra  
= i\bar{u}_{B_{f} } (p_{f}
) (A   + B \gamma _{5})u_{B_{i} } (p_{i} ),
\end{equation}
 where $u_{B_i} $ represent Dirac spinors for initial and final ($\frac{1}{2}^+$) baryons $B_i$ and $B_f$. $A$ and $B$ denotes  the parity-violating (PV) \textit{s}-wave and parity-conserving (PC) \textit{p}-wave amplitudes, respectively. 

The decay rate formula for $B_i\to B_{f}P$ process is given by
\ben
{\Gamma (B_i\to B_{f}P) =\frac{p_c }{8\pi } \frac{E_{f} +m_{f} }{m_{i} } \Big[
|A|^{2} +\frac{E_{f} -m_{f} }{E_{f} +m_{f} } (|B|^{2} )\Big ] }.
\en
Here $m_{i} $ and $m_{f} $ are the masses of the initial and final state baryons. The magnitude of the three-momentum $p_c$ of the final-state particles in the rest frame of $B_i$ is  
\ben p_{c } = \frac{1}{2m_i}\sqrt{[m_{i}^{2} -(m_{f} -m_{P} )^{2} ][m_{i}^{2} -(m_{f} +m_{P} )^{2} ]} ,\non \en
  where $m_{P} $ is the mass of emitted pseudoscalar meson, and  
\ben
E_{f} \pm m_{f}= \frac{(m_{i} \pm m_{f})^{2} -m_{P}^{2}}{2m_i}. \non
\en 
 The corresponding asymmetry parameter is given by
\ben
\alpha =\frac{2~\frac{p_c}{E_f + m_f }~\mbox{Re}[A*B]}{
(|A|^{2} + \frac{p_c^2}{(E_f + m_f )^2}|B|^{2} )}.\en
To estimate the decay rate and asymmetry parameters we require to calculate numerically the amplitudes, $A$ and $B$.

\section{Decay Amplitudes}
The hadronic matrix element for the $B_i \to B_f +P_k$ process can receive dominant contributions from factorization and pole processes, thus can be given as follows: 
\ben \langle B_{f} P_{k}\vert H_{W} \vert B_{i} \rangle \equiv \mathcal{A}_{Pole} + \mathcal{A}_{Fac.} ,  \en
where $\mathcal{A}_{Pole}$ and $\mathcal{A}_{Fac.}$ denotes pole and factorization amplitudes, respectively. The pole diagrams mainly involves the W-exchange process contributions that are evaluated using the pole model framework \cite{63}. In the pole model, the weak and strong vertices are
separated by introduction of a set of intermediate states into the decay process. It may also be noted that factorization may be considered as a correction to pole contributions where $t-$channel pole process is equivalent to the tree-level diagram i.e. factorizable process. The contribution of both pole and factorization processes can be summed up in terms of \textit{s}-wave (PV) and \textit{p}-wave (PC) amplitudes. We wish to point out that we have ignored the relative strong phases involved in the decay amplitudes in our calculation for being difficult to estimate in the present scenario, however, such phases can contribute to some of the \textit{CP}-violating asymmetries.

\subsection{Pole amplitudes}
The decay amplitude, $\mathcal{A}_{Pole}$, can be calculated from the reduced matrix element
\ben
\la B_f \vert H \vert B_i \ra = \bar{u}_{B_{i}} (A +\gamma_{5} B )u_{B_{f}},
\en
between two $\frac{1}{2}^+$ baryon states expressed in terms of PV and PC amplitudes, \textit{A} and \textit{B}, receptively. The baryonic decay in pole model involves hadronic intermediate state which first is produced in the strong process and then go through a weak transition to the final baryon. Thus, $A$ and $B$ can simply be expressed in terms of masses, strong couplings and weak matrix elements. The pole amplitude consisting of contributions of $s$ and $u$ channels for positive-parity intermediate baryon $(J^{P} =\frac{1}{2}^+ )$ poles are denoted by $A^{pole}$ and $B^{pole}$,
\ben
A^{pole} =\mathop{\Sigma }\limits_{n} \left[\frac{g^{B_{f} B_{n}}_{P_{k}}  b_{ni} }{m_{i} +m_{n} } +\frac{ g^{B_{n} B_{i} }_{P_{k} } b_{fn}}{m_{f} +m_{n} } \right],
\en
\ben
B^{pole} =\mathop{-\Sigma }\limits_{n} \left[\frac{g^{B_{f} B_{n} }_{P_{k} } a_{ni} }{m_{i} -m_{n} } +\frac{ g^{B_{n} B_{i}  } _{P_{k}}a_{fn}}{m_{f} -m_{n} } \right],
\en
where $g^{ij}_k $ is the strong meson-baryon coupling constants. The weak baryon-baryon matrix elements $a_{ij} $ and $b_{ij}$ are  defined as
\begin{equation}
\la B_{i} \vert H_{W} \vert B_{j} \ra = \bar{u}_{B_{i} } (a_{ij} +\gamma _{5} b_{ij} )u_{B_{j} } .
\end{equation}
As a preliminary study, we will restrict ourself to the contributions from parity-conserving amplitudes for the following reasons:
\begin{enumerate}
\item It is well known that the PV matrix element $b_{ij}$ vanishes in SU(3) flavor symmetry limit i.e. $\langle B_{f} P_{k}\vert H_{W}^{PV} \vert B_{i} \rangle =0$. Since for charmed baryon decays $b_{ij} \ll a_{ij} $,  the contributions of $\frac{1}{2}^+-$ poles are expected to be suppressed in \textit{s-}wave amplitudes and dominant in $p$-wave amplitudes. Moreover, presence of sum of the baryon masses in the denominator further suppresses their contributions. Thus, consideration of PC terms only turns out to be a good approximation for heavy baryon decays.
\item Estimation of $\frac{1}{2}^{-}-$pole terms is nontrivial task in the present scenario as it involves knowledge of strong coupling constants and weak metrics elements of $\frac{1}{2}^{-}-$ baryons. 
\item Furthermore, it has been argued by Fayyazuddin and Riazuddin \cite{37} that, \textit{in the leading nonrelativistic approximation}, one can ignore $J^P=\frac{1}{2}^-,~ \frac{3}{2}^-....$ and higher (orbital) resonances in order to connect them to relevant the ground-state ($s$-wave) wave function in the overlap integral to satisfy the normalization condition: thus, only the PC amplitude survives.

\end{enumerate}

\subsection{Weak transitions}

The flavor symmetric and quark model weak Hamiltonian \cite{41,67} involved in weak transitions for the quark-level process $q_{i}  +q_{j} \to q_{l}  +q_{m} $ is given by
\begin{equation} \label{wH}
H_{W} \cong V_{il} V_{jm}^* c_{-}(m_c)[\bar{B}^{[i,j]k} B_{[l,m]k} H_{[i,j]}^{[l,m]} ],
\end{equation}
here $c_{-} = c_1 + c_2$, and the antisymmetrization among the indices is represented by the brackets, [~,~]. The spurion transforms like $H_{[2,4]}^{[1,3]} $. Equation (\ref{wH}) can be written in terms of the weak amplitude, $a_{W}$, for CKM-favored and CKM-suppressed modes:
\begin{equation}
H_{W} \cong a_W[\bar{B}^{[i,j]k} B_{[l,m]k} H_{[i,j]}^{[l,m]} ].
\end{equation}
As discussed in the literature \citep{42,70,71}, a rough estimate of $a_W$ can be made based on symmetry arguments. However, SU(4) symmetry (being badly broken) ignores QCD enhancements due to hard gluon exchanges, contributing through $c_{-}$, at corresponding mass scales, that will affect the weak transition. 

To calculate numerical values of pole terms, the weak matrix element $\langle B_{f}\vert H_{W}^{PC} \vert B_{i} \rangle $ can be treated in leading nonrelativistic approximation \cite{37}. Moreover, decays of doubly heavy baryons involve heavy-to-heavy transitions: thus, the use of nonrelativistic approximation suits the present analysis. Following the analysis of Riazuddin and Fayyazuddin \cite{37}, we obtained the weak transition amplitudes for the charm baryons as a \textit{first approximation}, 
\ben \label{wA}
\mathcal{M}^{PC}= \frac{G_F}{\sqrt{2}}V_{du} V_{cs} \sum_{i > j}(\gamma^{-}_i\alpha^{+}_j+\alpha^{+}_i \gamma^{-}_j ) (1-\sigma_i\cdot \sigma_j),\en
where \textbf{S$_i=$}$\mathbf{\sigma}_i/2$ are Pauli spinors representing the spin of $i$th quark. The operators $\alpha^{+}_i$ and $\gamma^{-}_j$ convert $d \to u$ and  $c \to s$ , respectively \cite{76}. The weak Hamiltonian can be obtained by using Fourier transformation of (\ref{wA}),
\ben \label{wA2} H^{PC}_W= \frac{G_F}{\sqrt{2}}V_{du} V_{cs} \sum_{i\neq j}\alpha^{+}_i \gamma^{-}_j (1-\sigma_i\cdot \sigma_j)\delta^3(r),\en
 which gives the first estimate of the pole terms. The spatial baryon wave function overlap, $\delta^3(r) \equiv \la \psi_f|\delta^3(r)|\psi_i\ra $, is usually assumed to be flavor invariant such that
\ben \label{scale} \la \psi_f|\delta^3(r)|\psi_i\ra_{c} \approx \la \psi_f|\delta^3(r)|\psi_i\ra_{s}.
\en 
 The relation (\ref{scale}) connects nonleptonic charmed baryon decays with hyperon decays in SU(4) symmetry. However, the SU(4) being badly broken due to the large mass difference between $s$ and $c$ quarks should yield a larger mismatch between strange and charm baryon wave function overlaps. Several methods have been proposed in the literature to address this issue by the introduction of a correction factor based on different arguments (for a summary, see Ref. \cite{71}). In the present analysis, we follow our previous work \cite{76} by treating $|\psi(0)|^2$ (based of dimensionality argument) as a flavor-dependent quantity. It may be noted that a reliable estimate of baryon ground-state wave function at the origin (at charm mass scale) can be obtained from, precisely known, experimental masses of baryons using hyperfine splitting, which in turn yields 
\begin{equation}
\frac{m_{\Sigma _{c}} -m_{\Lambda _{c}}}{m_{\Sigma} -m_{\Lambda} } =\frac{\alpha _{s} (m_{c} )}{\alpha _{s} (m_{s} )} \frac{m_{s} (m_{c} -m_{u} )|\psi (0)|_{c}^{2} }{m_{c} (m_{s} -m_{u} )|\psi (0)|_{s}^{2} }.
\end{equation}
Thus, we get
\begin{equation}
\frac{|\psi (0)|_{c}^{2} }{|\psi (0)|_{s}^{2} } \approx 2.1,
\end{equation}
for $\frac{\alpha _{s} (m_{c} )}{\alpha _{s} (m_{s} )}\approx 0.53$ \cite{70,76}. Thus, the variation of flavor-dependent baryon spatial wave function overlap would lead to a substantial correction in branching ratios of doubly heavy baryons. The numerical results are discussed in Sec. IV.   
\subsection{Strong coupling constants}

In general, meson-baryon strong couplings are obtained from the SU(4)-invariant strong Hamiltonian. In the present work, we follow a relatively accurate method used by Khanna and Verma \cite{74} to calculate the baryon-baryon-pseudoscalar (${\rm BB^{'}P}$) couplings. We extend their analysis to include SU(4)-breaking effects by employing the null result of Coleman and Glashow for the tadpole-type symmetry-breaking. The SU(4)-broken (SB) baryon-meson strong couplings are calculated by
\ben
g^{BB^{'}}_{P}\mbox{(SB)} = \frac{M_B +M_B^{'}}{2M_N} (\sqrt{\frac{8}{3}}\frac{m_s-m_u}{m_c-m_u}) g^{BB^{'}}_{P}\mbox{(Sym)},
\en
where $g^{BB^{'}}_{P}$(Sym) is the value of SU(4) symmetric couplings \cite{74,76}. Effects of symmetry-breaking are such that it should yield larger values of strong couplings as compared to symmetric ones due to mass dependence, consequently, leading to larger pole contributions for heavy-baryon decays. The obtained absolute numerical values and expressions of relevant SB strong meson-baryon coupling constants are presented in Table \ref{t1}. The $g^{BB^{'}}_{P} (SB)$ are expressed in terms of $g_D (=8.4)$ and $g_F (=5.6)$ \cite{41,78}. 

\subsection{Factorization}

The factorizable decay amplitudes (ignoring the scale factors) can be expanded in terms of the following reduced matrix elements:
\begin{equation}
\mathcal{A}^{Fac}(B_i \to B_f +P_k)\equiv <P_{k} (q)|A_{\mu } |0>\, <B_{f} (p_{f} )|V^{\mu } +A^{\mu } |B_{i} (p_{i} )>.
\end{equation}
The baryon-baryon matrix elements of the weak currents can be expressed in terms of form factors $f_i$ and $g_i$ (as functions of $q^2$) \cite{62,63} as

\begin{equation} \label{10}
<B_{f} (p_{f} )|V_{\mu } |B_{i} (p_{i} )>\, =\bar{u}_{f} (p_{f} )[f_{1}
\gamma _{\mu } -\frac{f_{2} }{m_{i} } i\sigma _{\mu \nu } q^{\nu } +\frac{%
f_{3} }{m_{i} } q_{\mu } ]u_{i} (p_{i} ),
\end{equation}
and

\begin{equation}  \label{11}
<B_{f} (p_{f} )|A_{\mu } |B(p_{i} )>\, =\bar{u}_{f} (p_{f} )[g_{1} \gamma
_{\mu } \gamma _{5} -\frac{g_{2} }{m_{i} } i\sigma _{\mu \nu } q^{\nu }
\gamma _{5} +\frac{g_{3} }{m_{i} } q_{\mu } \gamma _{5} ]u_{i} (p_{i} ).
\end{equation}
The decay constant $f_{P} $ of the emitted pseudoscalar meson, $%
P_{k} $, is defined as
\begin{equation}
<P_{k} (q)|A_{\mu } |0> = if_{P} m_{P}.
\end{equation}
\newpage



\begin{table}[h]
\captionof{table} {Expressions of strong-coupling constants and their absolute numerical values.}
\label{t1}
\renewcommand{\arraystretch}{1.16}
\centering
\begin{textblock}{3}[0.5,0.5](2,2.70)
\begin{tabular}{|c|c|c|}\hline
\multicolumn{2}{|c|}{Strong Couplings } & Absolute values    \\
\multicolumn{2}{|c|}{$g^{BB^{'}}_{P}$ } & $g^{BB^{'}}_{P}  $(SB)\\ \hline
$g^{\xic \lb}_{D}$ &$(\frac{ g_D }{\sqrt{2}}+ \frac{g_F}{3\sqrt{2}})$& 3.30 \\
$g^{\xicd \lb}_{D}$& $\frac{\sqrt{3}} {\sqrt{2}}(g_D- g_F)$ &1.60\\
$g^{\xicc \sic}_{D}$& $-(g_D+ g_F)$ &11.00\\
$g^{\xic \si}_{D}$&$(\sqrt{3} g_D + \frac{g_F}{\sqrt{3}})$ &8.40\\
$g^{\xicd \si}_{D}$& $(- g_D  +g_F)$ &1.40 \\
$g^{\xicp \lbc}_{D}$&$(\sqrt{3} g_D - \frac{g_F}{\sqrt{3}})$&8.60 \\
$g^{\xic \si}_{D}$& $(\frac{\sqrt{3}} {\sqrt{2}}g_D+ \frac{g_F}{\sqrt{6}})$&5.90 \\
$g^{\xicd \si}_{D}$&$(-\frac{ g_D }{\sqrt{2}}+ \frac{g_F}{\sqrt{2}})$&0.01 \\
$g^{\xic \x}_{\ds}$&$-{\sqrt{3}}g_D - \frac{g_F}{\sqrt{3}})$&8.60 \\
$g^{\xic \si}_{\ds}$& $(-g_D + g_F)$&1.40 \\
$g^{\xic \lbc}_{K}$ &$(  {\sqrt{2}}g_D- 2 {\sqrt{2}}\frac{g_F} 3)$ &16.70 \\
$g^{\xicd \lbc}_{K}$ &${\sqrt{2}}\frac{g_F}{\sqrt{3}}$ &11.80  \\
$g^{\xic \xic}_{\pi}$&$( g_D - 2\frac{g_F}3)$  &12.30\\
$g^{\xicd \xic}_{\pi}$ &$-\frac{g_F}{\sqrt{3}} $& 8.70\\
$g^{\xicp \xicp}_{\pi}$ &$(-g_D+g_F)$&10.80 \\
$g^{\xicp \xicp}_{\eta}$& $0.12g_D+0.08g_F$&1.50\\
$g^{\xic \xic}_{\eta}$& $-0.96 g_F$&14.50 \\
$g^{\xicd \xic}_{\eta}$ & $ 0.80 (g_D-g_F)$& 8.40\\
$g^{\xic \xic}_{\etap}$& $ 1.70 g_D-1.1 g_F $& 21.20\\
$g^{\xicd \xic}_{\etap}$ &$ 0.27g_F $&4.00\\
$g^{\xicp \xicp}_{\etap}$&$ 0.60(g_D-g_F)$ &6.90 \\
$g^{\xicp \xic}_{\pi}$& $-\sqrt{2}  \frac{g_F} 3$&12.30 \\
$g^{\xic \xic}_{\pi}$&$(  {\sqrt{2}}g_D- 2 {\sqrt{2}}\frac{g_F} 3)$&17.40 \\
$g^{\xic \sicp}_{K}$&$2\frac{g_F}{\sqrt{3}}$&17.00\\
$g^{\xicd \xicd}_{\pi}$ &$g_D$&23.10 \\
$g^{\xicd \xicd}_{\eta}$ &$ 0.12 g_D $ &2.80 \\
$g^{\xicp \xicp}_{\eta}$ &$ 0.77(g_D-g_F)$&8.40 \\
$g^{\xicd \xicd}_{\etap}$ & $1.7g_D$ & 39.80 \\
$g^{\xic \xicd}_{\pi}$& $-\frac{\sqrt{2}} {\sqrt{3}}g_F$&12.30\\
$g^{\sic \lbc}_{\pi}$ & $-\frac{2g_F}{\sqrt{3}}$ & 16.33 \\
\hline
\end{tabular}

\end{textblock}
\end{table}
\begin{table}[h]
\renewcommand{\arraystretch}{1.16}
\centering
\begin{textblock}{3}[0.5,0.5](4.5,2.69)
\begin{tabular}{|c|c|c|}\hline
\multicolumn{2}{|c|}{Strong Couplings } & Absolute values    \\
\multicolumn{2}{|c|}{$g^{BB^{'}}_{P}$ } & $g^{BB^{'}}_{P}  $(SB)\\ \hline
$g^{\xicd \xicd}_{\pi}$& ${\sqrt{2}} g_D$&32.60\\
$g^{\xic \oc}_{K}$ &$-2 \frac{g_F} {\sqrt{3}}$&17.80\\
$g^{\oc \xicd}_{K}$& $2 g_D$&47.20\\
$g^{\xicc \xic}_{\ds}$ &$(-\sqrt{3} g_D + \frac{g_F}{\sqrt{3}})$& 8.86 \\
$g^{\xicc \xicd}_{\ds}$& $-(g_D+g_F)$ &11.16\\
$g^{\xicc \occ}_{K}$&$\sqrt{2} (g_D - g_F)$ &15.46\\
$g^{\sicp p}_{D}$& $\sqrt{2} (g_D - g_F)$ &1.72 \\
$g^{\xicc \lbc}_{D}$&$(-\sqrt{3} g_D + \frac{g_F}{\sqrt{3}})$&8.60 \\
$g^{\sicp \si}_{\ds}$& $\sqrt{2}(g_D-g_F)$&1.85 \\
$g^{\xicc \xicd}_{\ds}$&$-(g_D +g_F)$&11.16 \\
$g^{\sicp \lbc}_{\pi}$& $\frac{2g_F}{\sqrt{3}}$&16.33 \\
$g^{\xicc \xicp}_{\pi}$ &$\sqrt{2} (g_D - g_F)$ &15.31 \\
$g^{\sicp \xic}_{K}$ &$\frac{2g_F}{\sqrt{3}}$ & 16.96 \\
$g^{\sicp \sicp}_{\pi}$&$2 g_D$  &43.91\\
$g^{\xicc \xicc}_{\pi}$ &$(g_D-g_F) $&10.82 \\
$g^{\sicp \sicp}_{\eta}$ &$1.55g_D$&33.98 \\
$g^{\xicc \xicc}_{\eta}$& $0.77(g_D-g_F)$&8.38\\
$g^{\sicp \sicp}_{\etap}$& $1.27 g_D$&27.81 \\
$g^{\xicc \xicc}_{\etap}$ & $ 0.63 (g_D-g_F)$&6.86 \\
$g^{\sicp \xicd}_{K}$& $ 2 g_D $&45.02 \\
$g^{\lbc p(n)}_{D}$ &$ (\sqrt{3} g_D + \frac{g_F}{\sqrt{3}}) $&7.37\\
$g^{\sic p(n)}_{D}$&$ (-g_D+g_F)$ &1.22 \\
$g^{\xicp \sic}_{D}$& $-\sqrt{2} (g_D + g_F)$&15.47 \\
$g^{\lbc \lb}_{\ds}$&$\sqrt{2}(g_D+\frac{g_F}{3})$&6.35 \\
$g^{\xicp \sic}_{\ds}$&$(-\sqrt{3} g_D + \frac{g_F}{\sqrt{3}})$&8.87\\
$g^{\sic \si}_{\ds}$ &$\sqrt{2} (g_D - g_F)$ &1.86 \\
$g^{\lbc \lbc}_{\eta}$& $1.55 g_D-1.03g_F$&17.60\\
$g^{\lbc \si}_{\ds}$ &0&0 \\
$g^{\lbc \lbc}_{\pi}$ &0&0 \\
$g^{\sic \lbc}_{\eta}$ &0&0\\
&&\\
\hline
\end{tabular}
\end{textblock}
\end{table}
\newpage

The factorizable amplitudes could be simplified to
\begin{equation*}
A_{1}^{fac} =-\frac{G_{F} }{\sqrt{2} } F_{C} f_{P} c_{k} [(m_{i} -m_{f})f_{1}^{B_{i} ,B_{f} } (m_{P}^{2} )],
\end{equation*}
\begin{equation*}
B_{1}^{fac} =\frac{G_{F} }{\sqrt{2} } F_{C} f_{P} c_{k} [(m_{i} +m_{f})g_{1}^{B_{i} ,B_{f} } (m_{P}^{2} ) ],
\end{equation*}

where the factor $F_C$ is a product of appropriate CKM factors and Clebsch-Gordan (CG)
coefficients and $c_k$ are corresponding QCD coefficients.

We use the NRQM \cite{72}  and the HQET \cite{73} to calculate the baryon-baryon transition form factors $f_{i} $ and $g_{i} $. In the NRQM calculations, the form factors are calculated in the Breit frame and include several corrections like the hard-gluon QCD contributions, the $q^2$ dependence of the form factors, and the wave-function mismatch. Later, in the heavy-quark sector, a $1/m_{Q}$ correction to the baryon-baryon transition form factors was introduced within the heavy-quark symmetry constraints using HQET. The obtained transition form factors are given in Table \ref{t2}.

\begin{table}[h]
\renewcommand{\arraystretch}{1.15}
\captionof{table} {$\xicc$ and $\xicp$ transition form factors in NRQM \cite{72}and HQET\cite{73}.}
\label{t2}
\begin{tabular}{|c|c|c|c|}
\cline{1-4}
Transitions                           & Models   & \multicolumn{2}{c|}{Form Factors} \\ \cline{3-4}
                                           & \cite{72}\cite{73} & $f_1$ & $g_1$\\ \hline
$\xicc \to \lbc$ & NRQM  & $-0.35$                   & $-0.19$                 \\
                 & HQET  & $-0.59$               & $-0.27$                 \\ \hline
$\xicc \to \sicp$& NRQM  & $-0.39$               & $-0.96$             \\
                   & HQET  & $-0.54$                 & $-1.35$             \\ \hline
\multicolumn{1}{|c|}{$\xicc \to \sic^+$} & NRQM  & $-0.27$               & $-0.68$             \\
\multicolumn{1}{|c|}{}                   & HQET  & $-0.38$                 & $-0.95$             \\ \hline
\multicolumn{1}{|c|}{$\xicc \to \xic^+$} & NRQM  & $-0.57$               & $-0.24$             \\
\multicolumn{1}{|c|}{}                   & HQET  & $-0.74$                 & $-0.29$             \\ \hline
\multicolumn{1}{|c|}{$\xicc \to \xic^{'+}$}& NRQM  & $-0.37$               & $-0.78$             \\
\multicolumn{1}{|c|}{}                   & HQET  & $-0.43$                 & $-0.91$             \\ \hline
\multicolumn{1}{|c|}{$\xicp \to \lbc$} & NRQM  & $0.35$               & $0.19$             \\
\multicolumn{1}{|c|}{}                   & HQET  & $0.59$                 & $0.27$            \\ \hline
\multicolumn{1}{|c|}{$\xicp \to \sic^+$} & NRQM  & $-0.27$               & $-0.68$             \\
\multicolumn{1}{|c|}{}                   & HQET  & $-0.38$                 & $-0.95$             \\ \hline
\multicolumn{1}{|c|}{$\xicp \to \sic^0$} & NRQM  & $-0.39$               & $-0.96$             \\
\multicolumn{1}{|c|}{}                   & HQET  & $-0.54$                 & $-1.35$            \\ \hline
\multicolumn{1}{|c|}{$\xicp \to \xic^0$} & NRQM  & $-0.57$               & $-0.24$             \\
\multicolumn{1}{|c|}{}                   & HQET  & $-0.74$                 & $-0.29$             \\ \hline
\multicolumn{1}{|c|}{$\xicp \to \xic^{'0}$}& NRQM  & $-0.37$               & $-0.78$             \\
\multicolumn{1}{|c|}{}                     & HQET  & $-0.43$                 & $-0.91$             \\ \hline
\end{tabular}
\end{table}
We use the mixing scheme for $\eta$ and $\etap$ mesons:
\ben \etap (0.958) &=& \frac{1}{\sqrt{2} } (u\overline{u}+d\overline{d})\cos \phi_{P} +(s\overline{s})\sin \phi_{P}, \non
  \\ 
  \eta(0.547)\,&=&\frac{1}{\sqrt{2} } (u\overline{u}+d\overline{d})\sin \phi_{P} -(s\overline{s})\cos \phi _{P},
\en
where $\phi_P = \theta_{ideal}-\theta_P ^{phy}$ and $\theta_P ^{phy} = -15.4^{\circ}$ \cite{79}. The decay constants \cite{79,80} relevant for the present analysis are given as
\begin{align}
f_{\pi} & =131~{\rm MeV},~~ f_{\eta}=133~{\rm MeV},~~ f_{\etap}=126~{\rm MeV},~~ f_{K}=160~{\rm MeV},\non \\ \non
f_{D} & =207.4~{\rm MeV}~~ {\rm and}~~ f_{D_{s}}=255~~{\rm MeV}.
\end{align}

\section{NUMERICAL RESULTS AND DISCUSSIONS}

The preliminary results for the various decay channels of $\Xi_{cc}$ are obtained as a sum of the factorization and the pole contributions to different PV and PC amplitudes. As mentioned before, SU(4) symmetry-breaking could be substantially large thus: the use of exact SU(4) symmetry could be questioned. Therefore, we include SU(4)-breaking effects in evaluating strong coupling constants as well weak transitions. First, we evaluate the factorizable amplitudes using NRQM-and HQET-based form factors for CKM-favored, CKM-suppressed and CKM-doubly-suppressed modes as listed in columns 3 and 4 of Tables \ref{t3}-\ref{t7}. The flavor-independent pole amplitudes are calculated by using SU(4) broken strong coupling constants as shown in column 5 of Tables \ref{t3}-ref{t7}.
\begin{table}[h]
\renewcommand{\arraystretch}{1.15}
\captionof{table} {Decay amplitudes (in units of $\frac{G_{F}}{ \sqrt{2}}  V_{uq} V_{cq}^{*}$) for CKM-favored ($\Delta C =\Delta S= -1$) mode.}
\label{t3}
\begin{tabular}{|c|c|c|c|c|c|}
\cline{1-6}
Decays   & Models   & \multicolumn{2}{c|}{Factorization\footnote{$A$ and $B$ represent PV and PC amplitudes, respectively.}} & \multicolumn{2}{c|}{Pole Amplitude}  \\ \cline{3-6}
& \cite{72}\cite{73} & & & Flavor &  Flavor\\ 
&  & $A^{fac}$ &$B^{fac}$  & independent & dependent\\ \hline
  
$\xicc \to \si^+ D^+$ & NRQM  & $0$   & $0$  & 0.101 & 0.212  \\

                   & HQET  & $0$    & $0$  &      &    \\ \hline

$\xicc \to \xic^+ \pi^+$ & NRQM  & $0.110$ & $-0.250$  & 0.372                & 0.782 \\
                     & HQET  & $0.142$ & $-0.290$  &           & \\ \hline
$\xicc \to \sicp \bar{K}^0$ & NRQM  & $-0.042$ & $0.520$  & 0                & 0\\
                     & HQET  & $-0.060$ & $0.730$  &           & \\ \hline
$\xicc \to \xic^{'+} \pi^+$ & NRQM  & $0.064$ & $-0.800$  & 0                & 0\\
                    & HQET  & $0.076$   & $-0.930$  &           & \\ \hline
\end{tabular}
\end{table}

\begin{table}
\renewcommand{\arraystretch}{1.00}
\captionof{table} {Decay amplitudes (in units of $\frac{G_{F}}{ \sqrt{2}}  V_{uq} V_{cq}^{*}$) for the CKM-favored ($\Delta C =\Delta S= -1$) mode.}
\label{t4}
\begin{tabular}{|c|c|c|c|c|c|}
\cline{1-6}
Decays   & Models   & \multicolumn{2}{c|}{Factorization} & \multicolumn{2}{c|}{Pole Amplitude}  \\ \cline{3-6}
& \cite{72}\cite{73} & & & Flavor &  Flavor\\ 
&  & $A^{fac}$ &$B^{fac}$  & independent & dependent\\ \hline
\multicolumn{1}{|c|}{$\xicp \to \lb^0 D^+$} & NRQM  & 0 & 0 & 0.082 & 0.172  \\
\multicolumn{1}{|c|}{}                     & HQET  & 0 & 0                  & &   \\ \hline
\multicolumn{1}{|c|}{$\xicp \to \si^+ D^0$} & NRQM  & $0$               & $0$  & 0.119        &     0.249       \\
\multicolumn{1}{|c|}{}                     & HQET  & $0$                 & $0$  &      &         \\ \hline
\multicolumn{1}{|c|}{$\xicp \to \si^0 D^+$} & NRQM  & $0$               & $0$  & 0.156       &  0.327           \\
\multicolumn{1}{|c|}{}                     & HQET  & $0$                 & $0$  &      &         \\ \hline
\multicolumn{1}{|c|}{$\xicp \to \Xi^0 \ds^+$} & NRQM  & $0$               & $0$  & $-0.114$     &  $-0.239$             \\
\multicolumn{1}{|c|}{}                     & HQET  & $0$                 & $0$  &      &         \\ \hline
\multicolumn{1}{|c|}{$\xicp \to \lbc \bar{K}^0$} & NRQM  & $0.043$               & $-0.102$  & $-0.407$                &   $-0.854$ \\
\multicolumn{1}{|c|}{}                     & HQET  & $0.072$                 & $-0.144$  &         &      \\ \hline
\multicolumn{1}{|c|}{$\xicp \to \xic^+ \pi^0$} & NRQM  & $0$               & $0$  & $-0.562$            &   $ -1.179$    \\
\multicolumn{1}{|c|}{}                     & HQET  & $0$                 & $0$  &       &        \\ \hline
\multicolumn{1}{|c|}{$\xicp \to \xic^{'+} \pi^0$} & NRQM  & $0$               & $0$  & 0.211       &     0.444        \\
\multicolumn{1}{|c|}{}                     & HQET  & $0$                 & $0$  &       &      \\ \hline
\multicolumn{1}{|c|}{$\xicp \to \xic^+ \eta$} & NRQM  & $0$               & $0$  & 0.240           &     0.504    \\
\multicolumn{1}{|c|}{}                     & HQET  & $0$                 & $0$  &        &       \\ \hline
\multicolumn{1}{|c|}{$\xicp \to \xic^{'+} \eta$} & NRQM  & $0$               & $0$  & 0.353     &   0.741            \\
\multicolumn{1}{|c|}{}                     & HQET  & $0$                 & $0$  &       &        \\ \hline
\multicolumn{1}{|c|}{$\xicp \to \xic^+ \etap$} & NRQM  & $0$               & $0$  & $-0.349$           &    $-0.733$     \\
\multicolumn{1}{|c|}{}                     & HQET  & $0$                 & $0$  &        &       \\ \hline
\multicolumn{1}{|c|}{$\xicp \to \xic^{'+} \etap$} & NRQM  & $0$               & $0$  & $-0.097$        &      $-0.205 $     \\
\multicolumn{1}{|c|}{}                     & HQET  & $0$                 & $0$  &        &       \\ \hline
\multicolumn{1}{|c|}{$\xicp \to \xic^{0} \pi^+$} & NRQM  & $0.110$               & $-0.250$  & $-0.422$       &            $-0.887$ \\
\multicolumn{1}{|c|}{}                     & HQET  & $0.143$                 & $-0.290$  &       &        \\ \hline
\multicolumn{1}{|c|}{$\xicp \to \xic^{'0} \pi^+$} & NRQM  & $0.064$               & $-0.802$  & 0.299      &             0.628 \\
\multicolumn{1}{|c|}{}                     & HQET  & $0.080$                 & $-0.940$  &     &          \\ \hline

\multicolumn{1}{|c|}{$\xicp \to \sicp K^-$} & NRQM  & $0$               & $0$  & $-0.412$             &    $-0.866$   \\
\multicolumn{1}{|c|}{}                     & HQET  & $0$                 & $0$  &       &        \\ \hline
\multicolumn{1}{|c|}{$\xicp \to \sic^+ \bar{K}^0$} & NRQM  & $-0.030$               & $0.370$  & $-0.291$         &        $-0.612$ \\
\multicolumn{1}{|c|}{}                     & HQET  & $-0.042$                 & $0.515$  &      &         \\ \hline
\multicolumn{1}{|c|}{$\xicp \to \oc K^+$} & NRQM  & $0$               & $0$   & 0.433        &     0.909      \\
\multicolumn{1}{|c|}{}                     & HQET  & $0$                 & $0$  &       &        \\ \hline
\end{tabular}
\end{table}

\begin{table}[h]
\renewcommand{\arraystretch}{1.15}
\captionof{table} {Decay amplitudes (in units of $\frac{G_{F}}{ \sqrt{2}}  V_{uq} V_{cq}^{*}$)  for the CKM-suppressed ($\Delta C = -1, \Delta S= 0$) mode.}
\label{t5}
\begin{tabular}{|c|c|c|c|c|c|}
\cline{1-6}
Decays   & Models   & \multicolumn{2}{c|}{Factorization} & \multicolumn{2}{c|}{Pole Amplitude}  \\ \cline{3-6}
& \cite{72}\cite{73} & & & Flavor &  Flavor\\ 
&  & $A^{fac}$ &$B^{fac}$  & independent & dependent\\ \hline
\multicolumn{1}{|c|}{$\xicc \to p D^+$} & NRQM  & 0 & 0 & 0.087 &  0.182    \\
\multicolumn{1}{|c|}{}                     & HQET  & 0 & 0                  & &      \\ \hline
\multicolumn{1}{|c|}{$\xicc \to \si^+ \ds^+$} & NRQM  & $0$                   & $0$  & 0.099              &     0.207  \\
\multicolumn{1}{|c|}{}                     & HQET  & $0$                 & $0$  &               &      \\ \hline
\multicolumn{1}{|c|}{$\xicc \to \lbc \pi^+$} & NRQM  & $0.078$               & $-0.190$  & 0.322                 &     0.676 \\
\multicolumn{1}{|c|}{}                     & HQET  & $0.131$                 & $-0.270$  &           &       \\ \hline
\multicolumn{1}{|c|}{$\xicc \to \xic^+ K^+$} & NRQM  & $0.150$               & $-0.320$  & 0.354                &      0.743 \\
\multicolumn{1}{|c|}{}                     & HQET  & $0.190$                 & $-0.380$  &           &       \\ \hline
\multicolumn{1}{|c|}{$\xicc \to \xic^{'+} K^+$} & NRQM  & $0.090$               & $-1.060$  & 0                &      0 \\
\multicolumn{1}{|c|}{}                     & HQET  & $0.100$                 & $-1.230$  &        &          \\ \hline
\multicolumn{1}{|c|}{$\xicc \to \sicp \pi^0$} & NRQM  & $0.022$               & $-0.280$  & 0          &            0 \\
\multicolumn{1}{|c|}{}                     & HQET  & $0.030$                 & $-0.400$  &          &        \\ \hline
\multicolumn{1}{|c|}{$\xicc \to \sicp \eta$} & NRQM  & $0.042$               & $-0.530$  & 0            &          0 \\
\multicolumn{1}{|c|}{}                     & HQET  & $0.062$                 & $-0.730$  &          &        \\ \hline
\multicolumn{1}{|c|}{$\xicc \to \sicp \etap$} & NRQM  & $-0.017$               & $0.170$  & 0         &             0 \\
\multicolumn{1}{|c|}{}                     & HQET  & $-0.023$                 & $0.230$  &         &         \\ \hline
\multicolumn{1}{|c|}{$\xicc \to \sic^+ \pi^+$} & NRQM  & $0.050$               & $-0.690$  & 0          &            0 \\
\multicolumn{1}{|c|}{}                     & HQET  & $0.080$                 & $-0.960$  &          &        \\ \hline
\end{tabular}
\end{table}
Later, we introduce the flavor-dependent effects in weak transition amplitudes through hyperfine splitting. The variation of the spatial baryon wave function overlap, $\|\psi (0)\|^{2} $, with flavor results in larger pole contributions. The numerical values flavor-dependent pole  amplitudes of $\Xi_{cc}$ decays in CKM-favored, CKM-suppressed and CKM-doubly-suppressed modes are given in column 6 of Tables \ref{t3}-\ref{t7}. It can be clearly seen that the pole contributions are enhanced by a factor of $\sim 2$ due to flavor-dependent effects caused by SU(4) breaking. Moreover, the increment in pole amplitudes could be viewed as variation of scale (charm to strange) by 2. 

We wish to remark that a significant contribution to the parity-violating amplitudes may come from, $\frac{1}{2}^-$ , the lowest-lying negative-parity excited
baryons, however, the estimation of such terms is far from simple, as discussed in \cite{63,64,65,66,71}. In addition, symmetry-based attempts have also been made to estimate their contributions for singly charmed baryons. Such attempts required sufficient experimental information on decays which is not available at present for doubly heavy $\Xi_{cc}$ baryons. Therefore, we have only considered ground-state $\frac{1}{2}^+$ intermediate baryon pole terms as a first estimate of pole contributions. It may be noted that a large theoretical uncertainty in the lifetime of $\Xi_{cc}$ states could be seen as another source of uncertainty in the results. We use $\tau_{\xicc}=300$ fs and $\tau_{\xicp}=100$ fs \cite{58} to obtain the branching ratios in the present work. 

\begin{table}
\renewcommand{\arraystretch}{1.10}
\captionof{table} {Decay amplitudes (in units of $\frac{G_{F}}{ \sqrt{2}}  V_{uq} V_{cq}^{*}$)  for the CKM-suppressed ($\Delta C = -1, \Delta S= 0$) mode.}
\label{t6}
\begin{tabular}{|c|c|c|c|c|c|}
\cline{1-6}
Decays   & Models   & \multicolumn{2}{c|}{Factorization} & \multicolumn{2}{c|}{Pole Amplitude}  \\ \cline{3-6}
& \cite{72}\cite{73} & & & Flavor &  Flavor\\ 
&  & $A^{fac}$ &$B^{fac}$  & independent & dependent\\ \hline
\multicolumn{1}{|c|}{$\xicp \to p D^0$} & NRQM  & 0 & 0 & $-0.111$ & $-0.234$     \\
\multicolumn{1}{|c|}{}                     & HQET  & 0 & 0                  &       &      \\ \hline
\multicolumn{1}{|c|}{$\xicp \to n D^+$} & NRQM  & $0$               & $0$  & $0.198$      & 0.416                 \\
\multicolumn{1}{|c|}{}                     & HQET  & $0$                 & $0$  &      &            \\ \hline
\multicolumn{1}{|c|}{$\xicp \to \lb^0 \ds^+$} & NRQM  & $0$               & $0$  & 0.056     & 0.117                 \\
\multicolumn{1}{|c|}{}                     & HQET  & $0$                 & $0$  &      &            \\ \hline
\multicolumn{1}{|c|}{$\xicp \to \si^0 \ds^+$} & NRQM  & $0$                   & $0$  & 0.070     & 0.147               \\
\multicolumn{1}{|c|}{}                     & HQET  & $0$                 & $0$  &     &                \\ \hline
\multicolumn{1}{|c|}{$\xicp \to \lbc \pi^0$} & NRQM  & $-0.022$               & $0.054$  & $-0.228$    &      $-0.478$             \\
\multicolumn{1}{|c|}{}                     & HQET  & $-0.037$                 & $0.077$  &           &       \\ \hline
\multicolumn{1}{|c|}{$\xicp \to \lbc \eta$} & NRQM  & $-0.044$               & $0.010$  & $0.194$      &        $0.407$         \\
\multicolumn{1}{|c|}{}                     & HQET  & $-0.074$                 & $0.144$  &     &             \\ \hline
\multicolumn{1}{|c|}{$\xicp \to \lbc \etap$} & NRQM  & $-0.018$               & $0.034$  & $-0.159 $       & $-0.333$              \\
\multicolumn{1}{|c|}{}                     & HQET  & $-0.028$                 & $0.046$  &        &          \\ \hline
\multicolumn{1}{|c|}{$\xicp \to \xic^+ K^0$} & NRQM  & $0$               & $0$  &   $-0.710$          &    $-$1.49       \\
\multicolumn{1}{|c|}{}                     & HQET  & $0$                 & $0$  &       &           \\ \hline
\multicolumn{1}{|c|}{$\xicp \to \xic^{'+} K^0$} & NRQM  & $0$               & $0$  & $-0.249$          & $-0.523$            \\
\multicolumn{1}{|c|}{}                     & HQET  & $0$                 & $0$  &      &            \\ \hline
\multicolumn{1}{|c|}{$\xicp \to \xic^0 K^+$} & NRQM  & $0.150$               & $-0.330$  & $-$0.352      & $-$0.739                \\
\multicolumn{1}{|c|}{}                     & HQET  & $0.190$                 & $-0.380$  &      &            \\ \hline
\multicolumn{1}{|c|}{$\xicp \to \xic^{'0} K^+$} & NRQM  & $0.087$               & $-1.060$  & $-$0.249          &            $-$0.523 \\
\multicolumn{1}{|c|}{}                     & HQET  & $0.103$                 & $-1.230$  &        &          \\ \hline
\multicolumn{1}{|c|}{$\xicp \to \sicp \pi^-$} & NRQM  & $0$               & $0$  & $-0.343$         &    $-0.721$          \\
\multicolumn{1}{|c|}{}                     & HQET  & $0$                 & $0$  &       &           \\ \hline
\multicolumn{1}{|c|}{$\xicp \to \sic^+ \pi^0$} & NRQM  & $0.015$               & $-0.200$  & 0.343      &   0.721              \\
\multicolumn{1}{|c|}{}                     & HQET  & $0.022$                 & $-0.275$  &       &           \\ \hline
\multicolumn{1}{|c|}{$\xicp \to \sic^+ \eta$} & NRQM  & $0.030$               & $-0.370$   & 0      &     0            \\
\multicolumn{1}{|c|}{}                     & HQET  & $0.043$                 & $-0.520$  &        &          \\ \hline
\multicolumn{1}{|c|}{$\xicp \to \sic^+ \etap$} & NRQM  & $-0.011$               & $0.122$  & 0      &   0              \\
\multicolumn{1}{|c|}{}                     & HQET  & $-0.016$                 & $0.171$  &          &        \\ \hline
\multicolumn{1}{|c|}{$\xicp \to \sic^0 \pi^+$} & NRQM  & $0.076$               & $-0.971$  & 0.343             &   0.721       \\
\multicolumn{1}{|c|}{}                     & HQET  & $0.110$                 & $-1.360$  &          &        \\ \hline
\end{tabular}
\end{table}

\begin{table}
\renewcommand{\arraystretch}{1.15}
\captionof{table} {Decay amplitudes (in units of $\frac{G_{F}}{ \sqrt{2}}  V_{uq} V_{cq}^{*}$) for the CKM-doubly suppressed ($\Delta C =-\Delta S= -1$) mode.}
\label{t7}
\begin{tabular}{|c|c|c|c|c|c|}
\cline{1-6}
Decays   & Models   & \multicolumn{2}{c|}{Factorization} & \multicolumn{2}{c|}{Pole Amplitude}  \\ \cline{3-6}
& \cite{72}\cite{73} & & & Flavor &  Flavor\\ 
&  & $A^{fac}$ &$B^{fac}$  & independent & dependent\\ \hline
  
$\xicc \to p \ds^+$ & NRQM  & $0$   & $0$  & 0.085  &       0.178 \\

                   & HQET  & $0$    & $0$  &   &             \\ \hline

$\xicc \to \lbc K^+$ & NRQM  & $0.110$ & $-0.025$  & 0.308                 & 0.647     \\
                     & HQET  & $0.180$ & $-0.360$  &            &      \\ \hline
$\xicc \to \sicp K^0$ & NRQM  & $-0.042$ & $0.520$  & 0                 &   0   \\
                     & HQET  & $-0.059$ & $0.730$  &            &      \\ \hline
$\xicc \to \sic^+ K^+$ & NRQM  & $-0.004$ & $0.045$  & 0                 &  0    \\
                    & HQET  & $-0.005$   & $0.063$  &            &      \\ \hline
$\xicp \to n \ds^+$ & NRQM  & $0$   & $0$  & $-0.085$ &  $-0.178$ \\
                     & HQET  & $0$  & $0$  &           &       \\ \hline
$\xicp \to \lbc K^0$ & NRQM  & $0.043$  & $-0.102$  & 0.308 & 0.647     \\
                     & HQET  & $0.072$ & $-0.144$  &            &      \\ \hline
$\xicp \to \sic^+ K^0$ & NRQM  & $-0.030$ & $0.370$  & 0                 &  0    \\
                     & HQET  & $-0.042$ & $0.515$  &            &      \\ \hline
$\xicp \to \sic^0 K^+$ & NRQM  & $0.104$ & $-1.283$  & 0                 &  0    \\
                   & HQET  & $0.150$  & $-1.797$  &            &      \\ \hline
\end{tabular}
\end{table}

\begin{table}
\renewcommand{\arraystretch}{1.15}
\captionof{table} {Branching ratios for the CKM-favored ($\Delta C =\Delta S= -1$) mode with only pole contributions. The branching ratios for an arbitrary lifetime can be obtained by using $(\frac{\tau_{\xicc}}{300})\t \mathcal{B}(B_i\to B_fP)$ and $(\frac{\tau_{\xicp}}{100})\t \mathcal{B}(B_i\to B_fP)$.}
\label{t8}
\begin{tabular}{|c|c|c|}
\cline{1-3}
Decays      & \multicolumn{2}{c|}{Branching ratios} \\ \cline{2-3}
 &Flavor independent & Flavor dependent \\ \hline
$\xicc \to \si^+ D^+$  & $2.0\times10^{-3}$         &$8.9\times10^{-3}$ \\
\hline

\multicolumn{1}{|c|}{$\xicp \to \lb^0 D^+$}   & $5.3\times10^{-4}$ &$2.4\times10^{-3}$  \\
\hline
\multicolumn{1}{|c|}{$\xicp \to \si^+ D^0$} & $9.4\times10^{-4}$ & $4.2\times10^{-3}$ \\
\hline
\multicolumn{1}{|c|}{$\xicp \to \si^0 D^+$}     & $1.6\times10^{-3}$                  & $7.0\times10^{-3}$  \\
\hline
\multicolumn{1}{|c|}{$\xicp \to \Xi^0 \ds^+$}   & $4.1\times10^{-4}$ & $1.8\times10^{-3}$ \\ \hline
\multicolumn{1}{|c|}{$\xicp \to \xic^+ \pi^0$}   & $1.1\times10^{-2}$                                   & $5.0\times10^{-2}$ \\\hline
\multicolumn{1}{|c|}{$\xicp \to \xic^{'+} \pi^0$}   & $1.2\times10^{-3}$                             &    $5.4\times10^{-3}$ \\\hline
\multicolumn{1}{|c|}{$\xicp \to \xic^+ \eta$}   & $1.4\times10^{-3}$                              & $6.4\times10^{-3}$      \\\hline
\multicolumn{1}{|c|}{$\xicp \to \xic^{'+} \eta$}   & $2.2\times10^{-3}$                             & $9.5\times10^{-3}$   \\\hline
\multicolumn{1}{|c|}{$\xicp \to \xic^+ \etap$}   & $7.9\times10^{-4}$                             & $3.5\times10^{-3}$   \\\hline
\multicolumn{1}{|c|}{$\xicp \to \xic^{'+} \etap$}   & $1.8\times10^{-5}$                              &  $8.1\times10^{-5}$   \\\hline

\multicolumn{1}{|c|}{$\xicp \to \sicp K^-$}   & $4.8\times10^{-3}$                             &  $2.1\times10^{-2}$   \\\hline
\multicolumn{1}{|c|}{$\xicp \to \oc K^+$}   & $2.2\times10^{-3}$                           &  $1.0\times10^{-2}$        \\\hline
\end{tabular}
\end{table}

\begin{table}
\renewcommand{\arraystretch}{1.15}
\captionof{table} {Branching ratios for the CKM-suppressed ($\Delta C = -1, \Delta S= 0$) and CKM-doubly suppressed ($\Delta C =-\Delta S= -1$) modes with only pole contributions.}
\label{t9}
\begin{tabular}{|c|c|c|}
\cline{1-3}
Decays      & \multicolumn{2}{c|}{Branching ratios} \\ \cline{2-3}
 &Flavor independent & Flavor dependent \\ \hline
\multicolumn{3}{|l|}{($\Delta C = -1, \Delta S= 0$)}\\ \hline
\multicolumn{1}{|c|}{$\xicc \to p D^+$}   & $1.4\times10^{-4}$  & $6.0\times10^{-4}$  \\\hline
\multicolumn{1}{|c|}{$\xicc \to \si^+ \ds^+$} & $7.7\times10^{-5}$ & $3.4\times10^{-4}$  \\\hline
\multicolumn{1}{|c|}{$\xicp \to p D^0$} & $7.6\times10^{-5}$ &  $3.4\times10^{-4}$ \\\hline
\multicolumn{1}{|c|}{$\xicp \to n D^+$} & $2.4\times10^{-4}$ & $1.1\times10^{-3}$  \\\hline
\multicolumn{1}{|c|}{$\xicp \to \lb^0 \ds^+$}                & $1.0\times10^{-5}$               & $4.5\times10^{-5}$   \\\hline
\multicolumn{1}{|c|}{$\xicp \to \si^0 \ds^+$}  & $1.3\times10^{-5}$               & $5.6\times10^{-5}$   \\\hline
\multicolumn{1}{|c|}{$\xicp \to \xic^+ K^0$} & $7.0\times10^{-4}$                &  $3.1\times10^{-3}$  \\\hline
\multicolumn{1}{|c|}{$\xicp \to \xic^{'+} K^0$}               & $6.2\times10^{-5}$       &  $2.7\times10^{-3}$ \\\hline
\multicolumn{1}{|c|}{$\xicp \to \sicp \pi^-$}                & $2.3\times10^{-4}$               &    $1.0\times10^{-3}$  \\\hline
\multicolumn{3}{|l|}{($\Delta C =-\Delta S= -1$)}\\ \hline
  
$\xicc \to p \ds^+$    & $5.7\times10^{-6}$     & $2.5\times10^{-5}$  \\ \hline
$\xicp \to n \ds^+$  & $1.9\times10^{-6}$  &   $8.4\times10^{-6}$ \\ \hline

\end{tabular}
\end{table}

After adding factorizable and pole contributions, we calculate the branching ratios and asymmetry parameters for two-body weak decays of doubly heavy $\Xi_{cc}$ baryons for the flavor-independent and flavor-dependent cases. To emphasize the importance of the W-exchange contribution to $\Xi_{cc}$ decays, we present our predictions for the branching ratios of $\Xi_{cc}$ decays receiving contributions only from pole amplitudes in Tables \ref{t8} and \ref{t9}. The prediction for branching ratios receiving contributions from both the factorization and pole or factorization-only are given in Tables \ref{t10}-\ref{t12} for CKM-favored, CKM-suppressed and CKM-doubly suppressed, respectively. We draw the following observations:

\begin{enumerate}
\item As expected, a large number of the $\Xi_{cc}$ decay channels receive contributions from the W-exchange process. upon comparison with factorizable contributions, we find that the pole amplitudes are not only equipollent but also are dominant in several decays.
\item In the CKM-favored $(\Delta C = \Delta S = -1)$ decay mode, most of the decays come from the pole diagrams alone, and only two of the decay channels come from factorization. the rest of the decays receive dominant pole contributions except for $\xicp \to \Xi_c^{'0}\pi^+$. The order of the branching ratios for all the decays  range from $10^{-2}$ to $10^{-5}$ for flavor-independent case. While the inclusion of flavor-dependent effects enhances the pole contributions, consequently the branching ratios of dominant modes become  $\mathcal{O} (10^{-1})$ $ \sim \mathcal{O}( 10^{-3})$.
\item The pole and factorizable amplitudes can interfere constructively or destructively in decay modes with both, factorizable and pole, contributions. The pole and factorization amplitudes interfere constructively, in $\xicc \to \xic^+ \pi^+$, $\xicp \to \xic^{'0} \pi^+$ and $\xicp \to \sic^+ \bar{K}^0$ decay channels, however, these interfere destructively in $\xicp \to \lbc \bar{K}^0$ and $\xicp \to \xic^0 \pi^+$ decays. because of flavor dependence, branching ratios of the most dominant modes: $\mathcal{B}(\xicc \to \xic^+ \pi^+)$, $\mathcal{B}(\xicp \to \xic^{'0} \pi^+)$, and $\mathcal{B}(\xicp \to \xic^+ \pi^0)$ are  enhanced by an order of magnitude, and the last decay comes from W-exchange diagrams only. The large decay width of $\xicc \to \xic^+ \pi^+$ decay makes it the best candidate to look out for in experimental searches.
\item The factorization contributions obtained from NRQM and HQET differ owing to the difference in form factors. The results based on HQET, in general, have larger values.
\item In the CKM-suppressed $(\Delta C = -1, \Delta S = 0)$ decay mode, the most of the dominant decays receive contributions from both pole and decay amplitudes via their constructive inference. The flavor-dependent branching ratios of such decay channels are $\mathcal{O} (10^{-2})$ $ \sim \mathcal{O}( 10^{-3})$ with a few exceptions. However, the pole-only decays have branching ratios of $\mathcal{O} (10^{-3})$ $ \sim \mathcal{O}( 10^{-5})$. The most dominant decays in this mode are: $\xicc \to \lbc \pi^+$, $\xicc \to \xic^+ K^+$ and $\xicp \to \sic^0 \pi^+$.
\item The decays, $\xicp \to \xic^{0} K^+$ and $\xicp \to \xic^{'0} K^+$, present an interesting case of destructive interference between pole and factorization terms. It is worth noting that in $\xicp \to \xic^{0} K^+$ decay pole and factorization contributions to PC amplitudes are roughly comparable, while the PC factorization amplitude in $\xicp \to \xic^{'0}  K^+$ is predominant. Experimental searches for such decays will provide a useful test of the theory.
\item The decay channels in CKM-doubly suppressed $(\Delta C = \Delta S = -1)$ modes have branching ratios $\mathcal{O} (10^{-4})$ $ \sim \mathcal{O}( 10^{-6})$. Only two of the decays attain contributions from the pole alone. The decays having both pole and factorization contributions have larger branching ratios. It is interesting to note that decays with factorization-only contributions have branching ratios comparable to the decays with pole-only contributions.
\item  We wish to point out that the flavor-dependent results enhance the contribution of pole terms roughly by a factor of 4, consequently, giving larger branching ratios. Thus, results based on flavor dependence and flavor-independent analyses provide a useful domain for experimental searches.
\end{enumerate}

\begin{table}
\renewcommand{\arraystretch}{1.15}
\captionof{table} {Branching ratios for the CKM-favored ($\Delta C =\Delta S= -1$) mode including factorization and pole contributions.}
\label{t10}
\begin{tabular}{|c|c|c|c|c|c|}
\cline{1-6}
Decays   & Models   & \multicolumn{2}{c|}{Branching ratios}  &  \multicolumn{2}{c|}{ Asymmetries ($\alpha$)} \\\cline{3-6}
&\cite{72}\cite{73} &Flavor independent & Flavor dependent&Flavor independent & Flavor dependent \\ \hline
 
$\xicc \to \xic^+ \pi^+$ & NRQM  & $7.8\times10^{-2}$ & $15.1\times10^{-2}$ & $-0.997$   & $-0.856$                 \\
                     & HQET  & $10.9\times10^{-2}$ & $18.5\times10^{-2}$ & $-0.991$  & $-0.942$  \\ \hline
$\xicc \to \sicp \bar{K}^0$ & NRQM  & $2.8\times10^{-2}$ & - & $-0.760$ & -  \\
                     & HQET  & $5.5\times10^{-2}$ & & $-0.760$   & - \\ \hline
$\xicc \to \xic^{'+} \pi^+$ & NRQM  & $6.4\times10^{-2}$ &- & $-0.780$      & - \\
                    & HQET  & $8.8\times10^{-2}$  & & $-0.780$   & - \\ \hline

\multicolumn{1}{|c|}{$\xicp \to \lbc \bar{K}^0$} & NRQM  & $6.0\times10^{-3}$            &  $2.7\times10^{-2}$ & $0.927$   & $0.504$ \\
\multicolumn{1}{|c|}{}                     & HQET  & $8.3\times10^{-3}$             & $2.7\times10^{-2}$   & $0.964$            & $0.785$ \\ \hline

\multicolumn{1}{|c|}{$\xicp \to \xic^{0} \pi^+$} & NRQM  & $1.3\times10^{-2}$        &    $2.7\times10^{-2}$   & $0.552$                 & $0.996$ \\
\multicolumn{1}{|c|}{}                     & HQET  & $2.1\times10^{-2}$          &    $3.3\times10^{-2}$   & $0.341$              & $0.972$ \\ \hline
\multicolumn{1}{|c|}{$\xicp \to \xic^{'0} \pi^+$} & NRQM  & $3.3\times10^{-2}$      &     $5.9\times10^{-2}$    & $-0.653$                   & $-0.502$ \\
\multicolumn{1}{|c|}{}                     & HQET  & $4.7\times10^{-2}$          &   $7.2\times10^{-2}$    & $-0.647$ & $-0.535$             \\ \hline
\multicolumn{1}{|c|}{$\xicp \to \sic^+ \bar{K}^0$} & NRQM  & $1.3\times10^{-2}$        &  $2.8\times10^{-2}$     & $-0.483$                 & $-0.336$ \\
\multicolumn{1}{|c|}{}                     & HQET  & $2.0\times10^{-2}$          &   $3.8\times10^{-2}$    & $-0.543$   & $-0.404$            \\ \hline
\end{tabular}
\end{table}

\begin{table}
\renewcommand{\arraystretch}{1.10}
\captionof{table} {Branching ratios for the CKM-suppressed ($\Delta C = -1, \Delta S= 0$) mode including factorization and pole contributions.}
\label{t11}
\begin{tabular}{|c|c|c|c|c|c|}
\cline{1-6}
Decays   & Models   & \multicolumn{2}{c|}{Branching ratios}  &  \multicolumn{2}{c|}{ Asymmetries ($\alpha$)} \\\cline{3-6}
&\cite{72}\cite{73} &Flavor independent & Flavor dependent&Flavor independent & Flavor dependent \\ \hline

\multicolumn{1}{|c|}{$\xicc \to \lbc \pi^+$} & NRQM  & $3.2\times10^{-3}$         &   $7.4\times10^{-2}$    & $-0.930$& $-0.690$                    \\
\multicolumn{1}{|c|}{}                     & HQET  & $5.8\times10^{-3}$        &     $1.0\times10^{-2}$     & $-1.000$& $-0.890$              \\ \hline
\multicolumn{1}{|c|}{$\xicc \to \xic^+ K^+$} & NRQM  & $5.1\times10^{-3}$       &   $8.0\times10^{-3}$      & $-0.970$& $-0.980$                    \\
\multicolumn{1}{|c|}{}                     & HQET  & $7.6\times10^{-3}$        &   $1.1\times10^{-2}$       & $-0.920$& $-1.000$              \\ \hline
\multicolumn{1}{|c|}{$\xicc \to \xic^{'+} K^+$} & NRQM  & $4.4\times10^{-3}$       &    -    & $-0.850$                   & - \\
\multicolumn{1}{|c|}{}                     & HQET  & $6.0\times10^{-3}$        &        & $-0.850$  & - \\ \hline
\multicolumn{1}{|c|}{$\xicc \to \sicp \pi^0$} & NRQM  & $5.3\times10^{-4}$      &    -     & $-0.700$   & -                \\
\multicolumn{1}{|c|}{}                     & HQET  & $1.0\times10^{-3}$        &         & $-0.690$   & -           \\ \hline
\multicolumn{1}{|c|}{$\xicc \to \sicp \eta$} & NRQM  & $1.3\times10^{-3}$          &  -   & $-0.780$                   & - \\
\multicolumn{1}{|c|}{}                     & HQET  & $2.7\times10^{-3}$          &       & $-0.780$   & -           \\ \hline
\multicolumn{1}{|c|}{$\xicc \to \sicp \etap$} & NRQM  & $5.7\times10^{-5}$          &  -   & $-0.980$                   & - \\
\multicolumn{1}{|c|}{}                     & HQET  & $1.1\times10^{-4}$           &      & $-0.980$  & -            \\ \hline
\multicolumn{1}{|c|}{$\xicc \to \sic^+ \pi^+$} & NRQM  & $3.2\times10^{-3}$         &  -    & $-0.690$                   & - \\
\multicolumn{1}{|c|}{}                     & HQET  & $6.3\times10^{-3}$           &      & $-0.690$   & - \\ \hline
\multicolumn{1}{|c|}{$\xicp \to \lbc \pi^0$} & NRQM  & $2.5\times10^{-4}$      &   $8.3\times10^{-4}$       & $-0.625$                   & $-0.360$ \\
\multicolumn{1}{|c|}{}                     & HQET  & $3.4\times10^{-4}$        &     $9.5\times10^{-4}$     & $-0.840$  & $-0.548$          \\ \hline
\multicolumn{1}{|c|}{$\xicp \to \lbc \eta$} & NRQM  & $1.2\times10^{-4}$       &    $3.0\times10^{-4}$     & $0.734$                   & $0.935$ \\
\multicolumn{1}{|c|}{}                     & HQET  & $2.8\times10^{-4}$        &    $4.2\times10^{-4}$      & $0.277$            & $0.960$ \\ \hline
\multicolumn{1}{|c|}{$\xicp \to \lbc \etap$} & NRQM  & $4.6\times10^{-5}$      &      $1.4\times10^{-4}$   & $-0.829$                   & $-0.517$ \\
\multicolumn{1}{|c|}{}                     & HQET  & $7.0\times10^{-5}$         &    $1.7\times10^{-4}$     & $-0.983$   & $-0.747$           \\ \hline
\multicolumn{1}{|c|}{$\xicp \to \xic^0 K^+$} & NRQM  & $1.1\times10^{-3}$        &    $1.3\times10^{-3}$    & $0.061$                  & $0.778$ \\
\multicolumn{1}{|c|}{}                     & HQET  & $1.8\times10^{-3}$         &    $2.0\times10^{-3}$     & $-0.052$  & $0.578$            \\ \hline
\multicolumn{1}{|c|}{$\xicp \to \xic^{'0} K^+$} & NRQM  & $1.0\times10^{-3}$        &   $6.2\times10^{-4}$     & $-0.950$                  & $-1.000$ \\
\multicolumn{1}{|c|}{}                     & HQET  & $1.4\times10^{-3}$         &    $9.7\times10^{-4}$     & $-0.940$ & $-1.000$            \\ \hline
\multicolumn{1}{|c|}{$\xicp \to \sic^+ \pi^0$} & NRQM  & $5.8\times10^{-4}$         &   $1.7\times10^{-3}$    & $-0.290$                   & $-0.173$ \\
\multicolumn{1}{|c|}{}                     & HQET  & $7.7\times10^{-4}$         &     $2.0\times10^{-3}$    & $-0.351$  & $-0.222$            \\ \hline
\multicolumn{1}{|c|}{$\xicp \to \sic^+ \eta$} & NRQM  & $2.3\times10^{-4}$       &   -     & $-0.780$                   & -\\
\multicolumn{1}{|c|}{}                     & HQET  & $4.5\times10^{-4}$          &       & $-0.780$  & - \\ \hline
\multicolumn{1}{|c|}{$\xicp \to \sic^+ \etap$} & NRQM  & $1.0\times10^{-5}$        &    -   & $-0.980$                   & - \\
\multicolumn{1}{|c|}{}                     & HQET  & $1.9\times10^{-5}$        &         & $-0.980$    & -\\ \hline
\multicolumn{1}{|c|}{$\xicp \to \sic^0 \pi^+$} & NRQM  & $3.7\times10^{-3}$        &   $5.9\times10^{-3}$     & $-0.552$                  & $-0.443$ \\
\multicolumn{1}{|c|}{}                     & HQET  & $6.3\times10^{-3}$       &      $9.1\times10^{-3}$     & $-0.589$  & $-0.498$           \\ \hline

\end{tabular}
\end{table}
\begin{table}
\renewcommand{\arraystretch}{1.15}
\captionof{table} {Branching ratios for the CKM-doubly suppressed ($\Delta C =-\Delta S= -1$) mode including factorization and pole contributions.}
\label{t12}
\begin{tabular}{|c|c|c|c|c|c|}
\cline{1-6}
Decays   & Models   & \multicolumn{2}{c|}{Branching ratios}  &  \multicolumn{2}{c|}{ Asymmetries ($\alpha$)} \\\cline{3-6}
&\cite{72}\cite{73} &Flavor independent & Flavor dependent&Flavor independent & Flavor dependent \\ \hline
 
  $\xicc \to \lbc K^+$ & NRQM  & $2.1\times10^{-4}$ & $3.8\times10^{-4}$  & $-1.000$& $-0.860$    \\
                     & HQET  & $4.2\times10^{-4}$& $6.2\times10^{-4}$ & $-0.970$ & $-1.000$   \\ \hline
$\xicc \to \sicp K^0$ & NRQM  & $7.6\times10^{-5}$ &- & $-0.760$   & - \\
                     & HQET  & $1.5\times10^{-4}$ & & $-0.760$   & - \\ \hline
$\xicc \to \sic^+ K^+$ & NRQM  & $2.3\times10^{-4}$&- & $-0.760$& -    \\
                    & HQET  & $4.6\times10^{-4}$ &  & $-0.760$   & - \\ \hline
$\xicp \to \lbc K^0$ & NRQM  & $2.5\times10^{-5}$ & $7.2\times10^{-5}$ & $-0.802$ & $-0.510$   \\
                     & HQET  & $3.9\times10^{-5}$&$8.9\times10^{-5}$  & $-0.964$& $-0.731$    \\ \hline
$\xicp \to \sic^+ K^0$ & NRQM  & $1.3\times10^{-5}$ &- & $-0.760$                   & - \\
                     & HQET  & $2.5\times10^{-5}$ & & $-0.760$   & - \\ \hline
$\xicp \to \sic^0 K^+$ & NRQM  & $1.5\times10^{-4}$&- & $-0.760$                   & - \\
                   & HQET  & $3.0\times10^{-4}$&  & $-0.760$& -    \\ \hline
\end{tabular}
\end{table}

\par To compare our results with other works, we present corresponding decay modes in Table \ref{t13}. We first compare our results with some of the very recent analyses of nonleptonic decays $\xix$ baryons based on the factorization scheme \cite{58,59}. W. Wang \textit{et al.} \cite{58} have given an analysis of weak decays of doubly heavy baryons in the quark-diquark picture using the light front approach. Their branching ratios for dominant CKM-favored modes $\mathcal{B}(\xicc \to \xic^{(')+}\pi^+)$ and $\mathcal{B}(\xicp \to \xic^{(')0}\pi^+)$ are of the order of a few percent. The $\mathcal{B}(\xicc \to \xic^{'+}\pi^+)$ compares well with our result with no pole contribution (owing to a zero CG coefficient of baryon-baryon weak coupling for W-exchange pole terms\footnote{The weak coupling $a_{\xicp \Xi_c^{'+}}$  becomes zero following the operation of $(1-\sigma_i \cdot \sigma_j)$ on the wave function using (\ref{wA2}): for details, see Ref. \cite{37}.}). Despite the inclusion of dominant pole contributions and constructive interference between PC pole and factorization amplitudes, our result for the most dominant $\mathcal{B}(\xicc \to \xic^{+}\pi^+)$ is comparable to their result, i.e. $7.24 \%$. Thus, the major difference in results is due to the different form factors used in both the works. As mentioned before, $\xicp \to \xic^{0}\pi^+$ and $\xicp \to \xic^{'0}\pi^+$ represent peculiar cases of destructive and constructive interference between pole and factorization amplitudes, respectively. Therefore, the magnitude of the $\mathcal{B}(\xicp \to \xic^{0}\pi^+)$  in our case is smaller as compared to their branching $2.4 \%$ and vice versa for $\mathcal{B}(\xicp \to \xic^{'0}\pi^+)$. Similarly, for CKM-suppressed and CKM-doubly suppressed modes, branching ratios are of same order when compared with Ref. \cite{58}, i.e. $\mathcal{O}(10^{-3})$ and  $\mathcal{O}(10^{-4})$, respectively. In general, our results for branching ratios including both pole and factorization amplitudes are larger than their values as expected. The decay $\Xi_{cc}^{++} \to \Sigma_c^{++} \bar K^{*0}$ is first figured as a four-body process in Ref. \cite{60} which is predicted to be one of the most dominant modes. Thomas Gutsche \textit{et al.} \cite{59} have analyzed weak decay of $\xicc$ as decay chain  $\Xi_{cc}^{++} \to \Sigma_c^{++} (\to \Lambda_c^+ \pi^+)
+ \bar K^{*0} (\to K^-  \pi^+)$, which is expected to be experimentally favored due to the dominant branching ratios of the daughter decays. The $\xicc \to \sic^{++} \bar{K}^{(*)0}$ decays are studied using the factorization scheme in CCQM. The obtained branching ratio: $\mathcal{B}(\xicc \to \sic^{++} \bar{K}^{0})= 1.5\%$ at 300 fs, is of the same order when compared with our result. Other than the factorization scheme, the nonperturbative long-distance (W-exchange) contributions to $\x_{cc}$ decays have been calculated by Yu \textit{et al.} \cite{60}. The rescattering mechanism of FSIs, which has been ignored in the present work, is used to evaluate long-distance contributions. Authors have used the one-particle exchange method, where FSI is assumed to be dominated by rescattering of intermediate states \cite{81}. Thus, the amplitude is expressed in terms of strong coupling (of particles on mass shell) and form factor (for exchanged baryons that are off mass shell). Here, also, the branching ratios in case of the CKM-favored and CKM-suppressed modes for factorizable decay channels (see Table \ref{t13}) are of the same order as compared to our results. However, their branching ratios for (pole-only) $\xicp \to \sicp K^-$ and $\xicc \to p D^+$ decays are smaller by an order of magnitude as compared to our results for flavor-independent case. The difference in results may be attributed mainly to distinctive approaches. Although all the results compared here are based on different models/approaches, but they agree at least on the order of magnitude of the doubly charmed baryon decays. These results could be of great importance for experimentalists for future searches.

In the present work, we have ignored the \textit{CP} asymmetries as they have not yet been established in charmed baryon decays. However, like heavy-flavor mesons decays, the heavy-baryon decays are also prone to \textit{CP} violation. Even though it is well established that nonfactorizable diagrams like W-exchange/annihilation have a sizable impact on baryon decays, it would be a difficult task to establish \textit{CP} violation in charmed baryon decays as the \textit{CP} asymmetries originating from the Standard Model (SM) are very small or even zero \cite{82,83}. Moreover, the production of three-body final states with relatively larger branching ratios and many \textit{CP} observables will require large amount of experimental data. On the other hand, \textit{CP} asymmetries has already been probed in two-body $\lb_b$ decays \cite{78}. The theoretical investigation based on pQCD approach \cite{52} indicates the dominance of nonfactorizable contributions in addition to penguin amplitudes. Similar conclusions were made by theoretical estimates based on generalized factorization and symmetries  \cite{47, 48, 51, 84}. Obviously, measurements of the \textit{CP} asymmetries  provide a good tool to probe interference between the SM and new physics.
\begin{table}
\renewcommand{\arraystretch}{1.0}
\captionof{table} {Comparison of branching ratios with other works \footnote{The branching ratios are compared for the lifetime $\frac{\tau_{\Xi_{cc}^{++}}}{\tau_{\Xi_{cc}^{+}}}=3$,  and thus, for Ref. \cite{60}, we have used $\mathcal{R}_{\tau}=0.3$.}.}

\label{t13}
\begin{tabular}{|c|c|c|c|c|}
\cline{1-5}
Decays  & Models    & \multicolumn{3}{c|}{Branching ratios} \\ \cline{3-5}
&\cite{72}\cite{73} &Flavor independent & Flavor dependent & Other works \\ \hline
 \multicolumn{5}{|l|}{($\Delta C = \Delta S= -1$)}\\ \hline

$\xicc \to \xic^+ \pi^+$ & NRQM  & $7.8\times10^{-2}$ & $15.1\times10^{-2}$ & $7.24\times10^{-2}$                   \cite{58} \\
                     & HQET  & $10.9\times10^{-2}$ & $18.5\times10^{-2}$ & $3.4\times10^{-2}$ \cite{60}   \\ \hline

$\xicc \to \xic^{'+} \pi^+$ & NRQM  & $6.4\times10^{-2}$ &- & $5.08\times10^{-2}$ \cite{58}      \\
                    & HQET  & $8.8\times10^{-2}$  & &    \\ \hline

\multicolumn{1}{|c|}{$\xicp \to \xic^{0} \pi^+$} & NRQM  & $1.3\times10^{-2}$        &    $2.7\times10^{-2}$   & $2.40\times10^{-2}$\cite{58}                  \\
\multicolumn{1}{|c|}{}                     & HQET  & $2.1\times10^{-2}$          &    $3.3\times10^{-2}$   & $1.2\times10^{-2}$ \cite{60}             \\ \hline
\multicolumn{1}{|c|}{$\xicp \to \xic^{'0} \pi^+$} & NRQM  & $3.3\times10^{-2}$      &     $5.9\times10^{-2}$    & $1.68\times10^{-2}$ \cite{58}                    \\
\multicolumn{1}{|c|}{}                     & HQET  & $4.7\times10^{-2}$          &   $7.2\times10^{-2}$    &              \\ \hline
\multicolumn{1}{|c|}{$\xicp \to \sicp K^-$} & Pole only  & $4.8\times10^{-3}$                             &  $2.1\times10^{-2}$  & $4.8\times10^{-4}$\cite{60} \\\hline
\multicolumn{1}{|c|}{$\xicp \to \lb^0 D^+$}  & Pole only  & $5.3\times10^{-4}$ &$2.4\times10^{-3}$ & $2.4\times10^{-4}$ \cite{60}\\
\hline
\multicolumn{5}{|l|}{($\Delta C = -1, \Delta S= 0$)}\\ \hline
\multicolumn{1}{|c|}{$\xicc \to \lbc \pi^+$} & NRQM  & $3.2\times10^{-3}$         &   $7.4\times10^{-2}$    & $4.09\times10^{-3}$ \cite{58}\\
\multicolumn{1}{|c|}{}                     & HQET  & $5.8\times10^{-3}$        &     $1.0\times10^{-2}$     & $1.2\times10^{-3}$ \cite{60} \\ \hline
\multicolumn{1}{|c|}{$\xicc \to \xic^+ K^+$} & NRQM  & $5.1\times10^{-3}$       &   $8.0\times10^{-3}$      & $6.06\times10^{-3}$ \cite{58}                    \\
\multicolumn{1}{|c|}{}                     & HQET  & $7.6\times10^{-3}$        &   $1.1\times10^{-2}$       &  \\ \hline
\multicolumn{1}{|c|}{$\xicc \to \xic^{'+} K^+$} & NRQM  & $4.4\times10^{-3}$       &    -    & $3.48\times10^{-3}$ \cite{58} \\
\multicolumn{1}{|c|}{}                     & HQET  & $6.0\times10^{-3}$        &        &    \\ \hline
\multicolumn{1}{|c|}{$\xicc \to \sic^+ \pi^+$} & NRQM  & $3.2\times10^{-3}$         &   $7.4\times10^{-2}$    & $2.66\times10^{-3}$ \cite{58} \\
\multicolumn{1}{|c|}{}                     & HQET  & $6.0\times10^{-3}$        &        &    \\ \hline
\multicolumn{1}{|c|}{$\xicp \to \xic^0 K^+$} & NRQM  & $1.1\times10^{-3}$       &   $1.3\times10^{-3}$      &                    $2.00\times10^{-3}$ \cite{58}\\
\multicolumn{1}{|c|}{}                     & HQET  & $1.8\times10^{-3}$        &   $2.0\times10^{-3}$       &  \\ \hline
\multicolumn{1}{|c|}{$\xicp \to \xic^{'0} K^+$} & NRQM  & $1.0\times10^{-3}$       &    $6.2\times10^{-4}$    & $1.15\times10^{-3}$ \cite{58} \\
\multicolumn{1}{|c|}{}                     & HQET  & $1.4\times10^{-3}$        & $9.7\times10^{-4}$       &   \\ \hline
\multicolumn{1}{|c|}{$\xicp \to \sic^0 \pi^+$} & NRQM  & $3.7\times10^{-3}$         &   $5.9\times10^{-3}$    & $1.77\times10^{-3}$ \cite{58} \\
\multicolumn{1}{|c|}{}                     & HQET  & $6.3\times10^{-3}$        &  $9.1\times10^{-3}$      &    \\ \hline

\multicolumn{1}{|c|}{$\xicc \to p D^+$}  & Pole only & $1.4\times10^{-4}$  & $6.0\times10^{-4}$  &$4.8\times10^{-5}$\cite{60}\\\hline
\multicolumn{1}{|c|}{$\xicp \to p D^0$}& Pole only & $7.6\times10^{-5}$ &  $3.4\times10^{-4}$  & $1.2\times10^{-4}$\cite{60}\\\hline

\multicolumn{5}{|l|}{($\Delta C =-\Delta S= -1$)}\\ \hline

  $\xicc \to \lbc K^+$ & NRQM  & $2.1\times10^{-4}$ & $3.8\times10^{-4}$  & $3.60\times10^{-4}$ \cite{58}  \\
                     & HQET  & $4.2\times10^{-4}$ & $6.2\times10^{-4}$ &     \\ \hline

$\xicc \to \sic^+ K^+$ & NRQM  & $2.3\times10^{-4}$&- & $1.95\times10^{-4}$ \cite{58}   \\
                    & HQET  & $4.6\times10^{-4}$ &  &    \\ \hline

$\xicp \to \sic^0 K^+$ & NRQM  & $1.5\times10^{-4}$ &- & $1.30\times10^{-4}$ \cite{58}                  \\
& HQET  & $3.0\times10^{-4}$ &  &     \\ \hline

\end{tabular}
\end{table}
\section{SUMMARY}
The understanding of heavy-baryon decays is a long-standing problem as there does not exist a reliable approach for investigating the weak decays of heavy baryons as of yet. The dynamics of baryon decays, unlike meson decays, seems to get more complicated once they become heavier. Motivated by the recent observations, especially by LHCb, we have analyzed nonleptonic weak decays of doubly charmed baryons. The branching ratios of $\xix$ decays for CKM-favored and -suppressed modes are calculated using the factorization and pole model approaches. In the factorization scheme, we have obtained the form factors, $f_i$ and $g_i$, using nonrelativistic quark model \cite{72} and heavy quark effective theory \cite{73}. The nonfactorizable W-exchange diagrams, involving $\frac{1}{2}^+$ intermediate states, are calculated using the pole model approach . In the case of singly charmed baryon decays, it has been  well established that the W-exchange contributions are comparable to factorization amplitudes. Therefore, the purpose of the present work is to give first estimates of W-exchange terms in doubly charmed $\xix$ decays to get a more comprehensive picture. As mentioned before, there has been some recent analysis involving doubly heavy baryons based mostly on factorization contributions only. However, the importance of W-exchange terms has also been emphasized in such works. Furthermore, we include SU(4)-breaking effects in meson-baryon strong couplings as well as in weak amplitudes. The results for the two scenarios, namely, flavor-independent and flavor dependent have been presented. We summarize our observations as follows:
\begin{enumerate}
\item We find that W-exchange amplitude contributes to the majority of the $\xix$ decays. In contrast to factorization contributions, the W-exchange contributions are not only comparable but also dominant in many decay channels. Thus, W-exchange contributions in $\xix$ decays cannot be ignored. 
\item It is interesting to note that most of the CKM-favored decay channels receive contributions from W-exchange pole amplitudes only. The overall branching ratios in this mode range from $10^{-1} \sim 10^{-5}$, The $\mathcal{B}(\xicc \to \xic^+ \pi^+)$ is as high as $\mathcal{O}(10^{-1})$ in flavor-dependent case. Several decays in this node have branching ratios of the order of a few percent, which could be of experimental interest.
\item We have shown that the pole and factorization amplitudes,  depending on their signs, can interfere constructively and destructively. An experimental search of these decays could prove to be a useful test of theoretical models. 
\item In CKM-suppressed and CKM-doubly suppressed modes, the dominant decays receive contributions from factorization as well as pole amplitudes indicating the importance of W-exchange processes. The branching ratios of dominant decay channels in the CKM-suppressed mode are $\mathcal{O}(10^{-2})~\sim~ \mathcal{O}(10^{-3})$.         
\item The pole contributions are significantly enhanced due to the flavor-dependent factor. Thus, our results based on the NRQM and HQET picture alongside flavor-dependent W-exchange contributions provide a useful range to search for experimental evidence. 

\end{enumerate}

Experimental searches for heavy-baryon decays could help theorists understand the underlying dynamics of W-exchange processes in such decays. The importance of nonfactorizable contributions in \textit{CP} asymmetries in heavy-baryon decays could prove to be a challenge to the theory as well as experiment. New measurements on of doubly heavy baryons are in future plans of several ongoing experiments at Fermilab and CERN. We hope that our results could prove to be useful in experimental searches for new modes.

\end{document}